\documentclass[12pt]{article}
\usepackage{psfig,epsfig}
\def\bd{
\begin{document}} \def\ed{\end{document}}
\def\beq{\begin{equation}} \def\eeq{\end{equation}}
\def\bea{\begin{eqnarray}} \def\eea{\end{eqnarray}} \def\rt{{\rm t}}
\def\bcc{\begin{center}} \def\ecc{\end{center}}
\def\f{\left} \def\g{\right} \def\e{{\rm e}}
\def\p{\partial} \def\cov{{\rm cov}} \def\ch{{\rm ch}}
\def\vf{\varphi} \def\EE{e$^+$e$^-$}  \def\pt{p_{\rm t}}
\def\cl{\centerline} \def\fnsz{\footnotesize}
\def\ub{\underbar} \def\hs{\hskip} \def\vs{\vskip} \def\ni{\noindent}
\def\pa{\parindent} \def\ej{\vfill\eject} \def\hf{\hfill}
\def\la{\langle} \def\ra{\rangle} \def\r#1{$^{[#1]}$}
\def\qqg{q$\bar{\rm q}$g\hskip1pt} \def\epn{e$^+$e$^-$\hskip1pt }
\def\pt{{p_{\rm t}}} \def\vf{\varphi} \def\yct{y_{\rm cut}}
\def\NAtwo{{\sc na}{\footnotesize 22}}  \def\QGP{{\sc qgp}}
\def\NPM{{\sc npm}} \def\NFM{{\sc nfm}} \def\EFM{{\sc efm}}
\def\EPM{{\sc epm}}
\def\d{{\rm d}} \def\ebe{event-by-event} \def\JACEE{{\sc jacee}}
\bd

\cl{\Large Extracting Event Dynamics from Event-by-Event
Analysis$^\dag$}

\vskip1.5cm \cl{Fu Jinghua and Liu Lianshou}

\vskip0.5cm \cl{\small Institute of Particle Physics, Huazhong
Normal University, Wuhan 430079 China}

\vskip0.5cm \cl{\small fujh@iopp.ccnu.edu.cn;\quad
liuls@iopp.ccnu.edu.cn}

\vskip2.5cm

\cl{\large ABSTRACT}

 The problem of eliminating the statistical
fluctuations and extracting the event dynamics from event-by-event
analysis is discussed. New moments $G_p$ (for continuous
distribution), and $G_{q,p}$ (for anomalous distribution) are
proposed, which are experimentally measurable and can eliminate
the Poissonian type statistical fluctuations to recover the
dynamical moments $C_p$ and $C_{q,p}$. In this way, the dynamical
distribution of the event-averaged transverse momentum $\bar
{\pt}$ can be extracted, and the anomalous scaling of dynamical
distribution, if exists, can be recovered, through event-by-event
analysis of experimental data.

\vskip2cm

 {\bf PACS:} \ 13.85.Hd

\vskip1cm

{\bf Keywords:} \ Multiparticle production; Event dynamics;
Event-by-event

\hskip2cm analysis

\vfill

$\dag$ \ This work is supported in part by the NSFC under project
90103019.

\section{Introduction}

In the conventional investigation of high energy multiparticle
production, all the events in a collision process are taken as a
whole and the distributions, fluctuations and correlations inside
the sample are studied without distinguishing the individual
events. This kind of study is usually referred to as inclusive
one. In the 80's -- 90's of last century, motivated by the
experimental observation~\cite{JACEE}~\cite{NA22fluc} and
theoretical perspective~\cite{QGP}, people started to carry on the
study event by
event~\cite{DDK}~\cite{NA49}\cite{PHENIX}\cite{CERES}, with the
aim of exploring the possible existence of new physics, such as
quark-gluon plasma (\QGP)~\cite{Harris} or non-linear
dynamics~\cite{BP}.

One of the differences in these two kinds of study is that, the
inclusive dynamics can readily be extracted from the experimental
measurement through averaging over a large number of events, while
the extracting of event dynamics is complicated due to the
statistical fluctuations coming from the limited number of
particles in a single event.

For concreteness, let us consider the \ebe\ fluctuation of the
transverse momentum distribution $p(\pt)$, a knowledge of which is
important for the understanding of the basic collision
dynamics~\cite{pt}. Usually, it is convenient to ``coarse-grain''
$p(\pt)$, {\it i.e.} to divide the phase space region $\Delta$ of
$\pt$ into $M$
bins and integrate $p(\pt)$ over $\pt$ in the $m$th bin $\delta_m$ 
\beq p_m=\int_{\delta_m} p(\pt) \d \pt, \qquad (m=1,2,\dots, M).  \eeq 
The set 
\beq p_1, p_2, \dots, p_M \eeq 
is the ``coarse-grained'' distribution~\cite{BP}. When $M\to
\infty$ ($\delta_m\to 0$) it recovers the original distribution
$p(\pt)$~\cite{note1}. If $p(\pt)$ is continuous this process is
convergent and a good approximation for $p(\pt)$ could be obtained
for not very large $M$.

The realization of $p(\pt)$ in experiment is the distribution of
the total number $N$ of particles in the $\pt$ region $\Delta$,
and the expression 
\beq q_m = {N_m}/{N} \qquad (m=1,2,\dots, M) \eeq 
is used to evaluate $p_m$, where $N_m$ is the number of particle
falling into the $m$th bin. This is, however, exact only when
$N\to\infty$. Just at this point appears the difference between
the inclusive and \ebe\ studies.

In an inclusive study, what is under consideration is the
transverse momentum distribution $p_{\rm incl}(\pt)$ in the {\it
event sample}. In this case, the number $N$ in Eq.(3) is the total
number of particles in the whole sample and could be made
arbitrarily large through increasing the number of events in the
experiment and thus the effect of statistical fluctuation can be
gotten rid of through averaging over a sufficiently large event
sample. On the contrary, in an \ebe\ study the distribution
$p(\pt)$ in consideration is the transverse momentum distribution
in a {\it single event} and the number $N$ in Eq.(3) is the number
$n$ of particles in the event, which is limited by energy
conservation, and the statistical fluctuation inevitably comes in.

Various methods have been proposed to eliminate the influence of
statistical fluctuation and evaluate the dynamical ones. Most of
them are based on the comparison of the measured fluctuation with
the expectation of statistically independent particle emission.
For example, in Ref.~\cite{PHENIX} the results from mixed events
are considered as the baseline for the random distribution and the
difference in the fluctuation from a random distribution defined
as 
\beq d=\omega_{\rm data}-\omega_{\rm baseline} \eeq  
is taken as a measure of the dynamical fluctuation. In Eq.(4) 
\beq \omega=\frac{\sqrt{\la \bar{\pt}^2\ra-\la \bar{\pt}\ra^2}}
{\la\bar{\pt}\ra}=\frac{\sqrt{\sigma^2_{\bar{\pt}}}}{\la\bar{\pt}\ra}, \eeq 
$\bar{\pt}$ is the mean transverse momentum in a single event and
$\la\cdots\ra$ denotes the average over event sample.

Alternatively, in Ref.~\cite{Voloshin} the statistical variance of
event mean $\pt$, under the assumption of independent particle
production, is estimated as 
\beq \sigma^2_{\bar{\pt}{\rm stat}}=\frac{\sigma^2_{\pt{\rm
incl}}}{\la n\ra}, \eeq 
and the difference between the variances of $\bar{\pt}$ obtained
from data and ``stat'' is taken as the dynamical variance 
\beq \sigma^2_{\bar{\pt}{\rm dynam}}=\sigma^2_{\bar{\pt}{\rm
data}}-\sigma^2_{\bar{\pt}{\rm stat}}=\sigma^2_{\bar{\pt}{\rm
data}}-\frac{\sigma^2_{\pt{\rm incl}}}{\la n\ra}. \eeq 

A widely used measure for the non-statistical mean $\pt$
fluctuation is the $\Phi_{\pt}$ proposed in Ref.~\cite{marek} 
\beq\Phi_{\pt} \equiv \sqrt{\la Z^2\ra/\la n \ra}
-\sqrt{\la{z^2}\ra}, \eeq 
where $z$ and $Z$ are defined as $z\equiv \pt-\la \pt\ra$ for each
particle and $Z\equiv \sum_{i=1}^n z_i = n(\bar{\pt}-\la\pt\ra)$
for each event, respectively~\cite{note2}. The second term of the
r.h.s. of Eq.(8) is the square root of the inclusive variance
$\sigma^2_{\pt{\rm incl}}=\la(\pt-\la\pt\ra)^2\ra$. Assuming that
the multiplicity fluctuation is uncorrelated with the $\pt$
fluctuation, we get from Eq.(8) 
\beq \Phi_{\pt}=\sqrt{\la n^2\ra/\la n \ra} \sigma_{\bar{\pt}{\rm
data}}-\sigma_{\pt{\rm incl}}. \eeq 
This equation is evidently similar to Eq.(7), both have the same
structure as Eq.(4), being based on a subtraction procedure, {\it
i.e.} to subtract the variance of $\bar{\pt}$ or a quantity
related to it, that will be expected from pure statistical
fluctuation, from the same quantity obtained in experiment. These
measures will, of course, vanish for a pure statistical system,
and a non-vanishing value of them will indicate the existence of
dynamical effect. Therefore, the measures based on the subtraction
procedure, as those listed above, will at least qualitatively
measures the effect of dynamical fluctuation.

The aim of the present paper is to develop a systematic method,
which is able to eliminate the statistical fluctuations directly
from the experimental \ebe\ analysis and extract quantitatively
the dynamical $\bar\pt$ moment of any positive integer order,
under the assumption that the statistical fluctuations are of the
Poisson type, i.e. due to uncorrelated random particle emission.

In Section II a theorem will be proved which is the basis of the
elimination of Poissonian statistical fluctuations. In Section III
the theorem is applied to the \ebe\ analysis of transverse
momentum distribution. In Section IV the \ebe\ fluctuation of the
non-linear fractal property will be discussed and the proposed
method will be applied to this case to extract the dynamical
fluctuation of fractal property. Section V is the conclusions.

\section{The elimination of Poissonian statistical fluctuation}

Divide the transverse momentum region $\Delta$ into $M$ bins. Let
$p_m$ be the event dynamical probability of $\pt$ in the $m$th
bin, {\it cf.} Eq.(1), and $n_m$ the number of particle in the
event with $\pt$ lying in the $m$th bin, then we have the theorem:
\beq \f\la \sum_{m=1}^M f_m p_m^p\g\ra = \f\la \sum_{m=1}^M
f_m \frac{n_m(n_m-1)\cdots(n_m-p+1)}{\la n\ra^p}\g\ra,  \eeq 
where $f_m$ is an arbitrary variable depending on $m$.

Before going on to prove this theorem, let us notice that the
symbol $\la\cdots\ra$ in the two sides of Eq.(10), despite of both
being the average over event sample, have different meanings. In
the l.h.s. it is simply the average over dynamical probability
distribution, while in the r.h.s. it includes also the average
over Poisson distribution of particle number. 
\beq \f\la \sum_{m=1}^M f_m p_m^p\g\ra = \int \sum_{m=1}^M f_m
p_m^p P(p_m) \d p_m, \eeq 
{\Large $ \f\la \sum_{m=1}^M f_m
\frac{n_m(n_m-1)\cdots(n_m-p+1)}{\la n\ra^p}\g\ra $} 
\beq = \int\sum_{m=1}^M \sum_{n_m=0}^\infty f_m
\frac{n_m(n_m-1)\cdots(n_m-p+1)}{\la n\ra^p}\cdot
\frac{\e^{-p_m\la n\ra
}}{n_m!} (p_m\la n\ra)^{n_m} P(p_m) \d p_m,  \eeq 
where $P(p_m)$ is the dynamical probability distribution of $p_m$
in the event space, $p_m\la n\ra=\la n_m\ra$ is the average
multiplicity in  the $m$th bin. Using the normalization condition
of Poisson distribution, it is easy to see that the r.h.s. of the
above two equations are equal and Eq.(10) follows.

Note that a simplified version of the above theorem with $f_m=1$
in Eq.(10) has been proved in Ref.~\cite{BP} and has been widely
used in the intermittency study~\cite{DDK}. Eq.(10) is a more
general theorem with a factor $f_m$ included, which is essential
in the application of this theorem to the elimination of
statistical fluctuation in \ebe\ analysis.

\section{The elimination of statistical fluctuation in \ebe\
transverse momentum fluctuation}

Let $\bar {\pt}$ and $\bar {\pt}_{\rm exp}$ be, respectively, the
dynamical and experimentally measured values of the event-averaged
$\pt$,  
\beq \bar {\pt}= \int_\Delta \pt p(\pt) \d \pt = \sum_{m=1}^M
(\pt)_m p_m, \eeq 
\beq \bar {\pt}_{\rm exp}  = \sum_{m=1}^M (\pt)_m q_m=\sum_{m=1}^M
(\pt)_m
\frac{n_m}{n}, \eeq 
where $(\pt)_m$ is the $\pt$ value in the $m$th bin. The event-space
moments of $\bar {\pt}$ and $\bar {\pt}_{\rm exp} $ are 
\beq C_p(\bar {\pt}) = \la \bar{\pt} ^p\ra = \f\la \f(\sum_{m=1}^M
(\pt)_mp_m\g)^p \g\ra,
\eeq 
\beq C^{\rm exp}_p(\bar{\pt}) = \la (\bar{\pt}_{\rm exp})^p\ra
= \f\la \f(\sum_{m=1}^M (\pt)_m\frac{n_m}{n}\g)^p  \g\ra, \eeq 
respectively.

Let us first consider the elimination of statistical fluctuations
in the second and third order moments. When $p=2,3$ we have  
 \bea C_2(\bar {\pt})\nonumber &=& \f\la \f(\sum_{m=1}^M (\pt)_mp_m\g)^2 \g\ra \\ 
&=&\f\la \sum_{m=1}^M (\pt)_m^2p_m^2 \g\ra 
+ \f\la \sum_{m\neq m'}^M (\pt)_m(\pt)_{m'}p_mp_{m'} \g\ra, \eea
\bea C_3(\bar {\pt}) \nonumber &=& \f\la \f(\sum_{m=1}^M (\pt)_mp_m\g)^3 \g\ra \\
\nonumber 
&=&\f\la \sum_{m=1}^M (\pt)_m^3p_m^3 \g\ra 
+ 3\f\la \sum_{m\neq m'}^M (\pt)_m (\pt)^2_{m'}p_mp^2_{m'} \g\ra \\ 
&+& \f\la \sum_{m\neq m'\neq m''}^M (\pt)_m (\pt)_{m'}
(\pt)_{m''}p_mp_{m'}p_{m''}
\g\ra . \eea
Define  
\bea G_2(\bar {\pt}) &=& \f\la \sum_{m=1}^M (\pt)_m^2
\frac{n_m(n_m-1)}{\la n\ra^2}
\g\ra 
+ \f\la \sum_{m\neq m'}^M (\pt)_m (\pt)_{m'}\frac{n_mn_{m'}}{\la n\ra^2} \g\ra, \\
 G_3(\bar {\pt}) \nonumber &=& \f\la \sum_{m=1}^M (\pt)_m^3
\frac{n_m(n_m-1)(n_m-2)}{\la n\ra^3} \g\ra 
+ 3\f\la \sum_{m\neq m'}^M (\pt)_m
(\pt)^2_{m'}\frac{n_mn_{m'}(n_{m'}-1)}{\la n\ra^3}
\g\ra \\ 
&+& \f\la \sum_{m\neq m'\neq m''}^M (\pt)_m (\pt)_{m'}
(\pt)_{m''}\frac{n_mn_{m'}n_{m''}}{\la n\ra^3}
\g\ra . \eea   
It is easy to see, using Eq.(10), that $G_2(\bar {\pt})=C_2(\bar
{\pt})$, $G_3(\bar {\pt})=C_3(\bar {\pt})$, provided the
statistical fluctuations are Poissonian. Thus the dynamical
moments $C_2, C_3$ can be extracted from the experimental
measurement by using $G_2, G_3$ and the statistical fluctuations
have been eliminated. The elimination of statistical fluctuations
in higher order moments can be proceeded in a similar manner.

\begin{center}
 \begin{figure}
 \includegraphics[width=12cm]{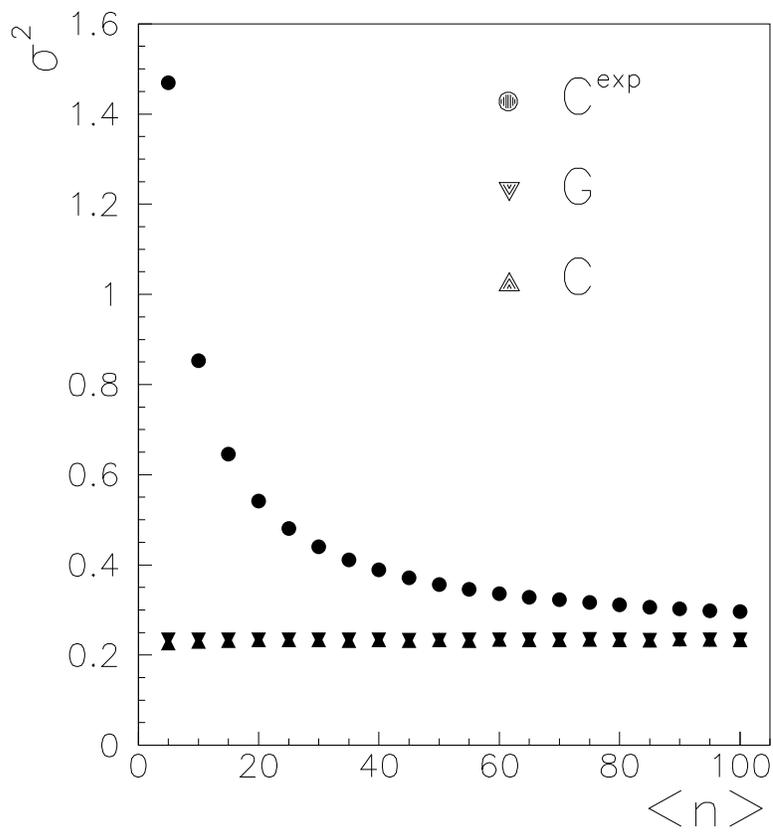}
 \caption{\label{fig:xbar} Variance $\sigma^2_{\bar{\pt}}$ of $\bar{\pt}$
 distribution}
 \end{figure}
 \end{center}

 Let us demonstrate the above results with a toy-model Monte
Carlo
simulation. In this model the distribution of $\pt$ is taken as 
\beq p(\pt) = \frac{4}{a^2} \pt \e^{-2\pt/a}, \eeq 
with a Gaussian distributed parameter $a$ ($\sigma^2(a)=0.24$). In
total 1,000,000 events have been generated. The resulting variance
$\sigma^2_{\bar{\pt}}=\la {\bar {\pt}}^2 \ra - \la \bar {\pt}\ra
^2$ for $C$, $C^{\rm exp}$ and $G$ are plotted in Fig.1 as upward
triangles, full circles and downward triangles, respectively. It
can be seen from the figure that the width
$\sqrt{\sigma^2_{\bar{\pt}}}$ of experimentally measured $\bar
{\pt}$ distribution is wider than that of the dynamical
distribution, especially when the average multiplicity is low,
while the width calculated from $G$ coincides with the dynamical
width, $\sigma^{2{\rm G}}_{\bar {\pt}} =\sigma^{2{\rm C}}_{\bar
{\pt}}$, where $\sigma^{2{\rm G}}_{\bar {\pt}} = G_2(\bar
{\pt})-\f(G_1(\bar {\pt})\g)^2$ and $G_1(\bar {\pt})=\la\sum_m
(\pt)_mn_m\ra/\la n\ra$. Therefore, using $G_p$ the Poissonian
statistical fluctuations are eliminated thoroughly and the event
dynamics is successfully extracted.

\section{The elimination of statistical fluctuation in \ebe\
analysis of fractal property}

In the above discussion the dynamical event-distributions of phase
space variables ------ rapidity $y$, transverse momentum $\pt$,
azimuthal angle $\vf$
------ are implicitly assumed to be
continuous functions, {\it i.e.} fulfil the condition 
\beq \delta x \to 0 \quad \Longrightarrow \quad \delta p(x)\to
0,\qquad x=y, \pt, \vf, \eeq 
which means that when the bin size decreases the probabilities in
neighboring bins tend to be equal to each other.

However, there is evidences showing that this may not be true in
some cases. The first experimental evidence is from a \JACEE\
event in 1983\cite{JACEE}, in which the total multiplicity is
about one thousand and the multiplicity fluctuations in a small
rapidity bin are still 2 $\times$ the average. Similar phenomena
have also been observed afterwards in accelerator experiments with
local fluctuations up to 60 $\times$ the average~\cite{NA22fluc}.
Obviously, this kind of fluctuations is out of the usual
statistical ones, indicating the existence of self-similar fractal
property in phase space distribution.

A characteristic phenomenon of self-similar fractal is the
anomalous scaling of the normalized probability moments (\NPM): 
\beq C_q(M) \equiv M^{-1}\f\la\sum_{m=1}^M (Mp_m)^q \g\ra
\propto M^{\phi_q}.\eeq
Such an anomalous scaling property can be demonstrated using a toy
model for self-similar fractal ------ random cascading $\alpha$
model~\cite{BP}\cite{KXTB}. This model describes each
multiparticle event as a series of steps, in which the initial
phase space region $\Delta$ is repeatedly partitioned into
$\lambda = 2$ parts. After $\nu$ steps we get $M = 2^\nu$
sub-cells of size $\delta =\Delta/M$. At each step $\nu$ the
probabilty  in each of the two parts is obtained by multiplication
of the probability in the step $\nu-1$ by a particular value of
the random variable $\omega^{(\nu)}_{j_\nu}$, where $j_\nu$ is the
position of a sub-cell at the $\nu$th step ($1\leq j_\nu\leq
2^\nu$). The elementary fluctuation probability $\omega$ can be
chosen in
various ways. The simplest way is to choose it as~\cite{KXTB} 
\beq \omega^{(\nu)}_{2j-1}=\frac{1}{2}(1+\alpha r), \quad
\omega^{(\nu)}_{2j}=\frac{1}{2}(1-\alpha r), \eeq 
where $r$ is a uniformly-distributed random number in the interval
[$-1,1$], $j$ is an integer ($1\leq j\leq 2^{\nu-1}$), $\alpha$ is
a characteristic parameter of the model taking value in the
interval [0,1].

In this model, after $\nu$ steps of partition, the
\NPM\ defined in Eq.(23) becomes 
\beq C_q(M)=M^q\la \omega^q(1)\cdots\omega^q(\nu)\ra. \eeq 
Since $\la \omega\ra = 1/2$, we have the anomalous scaling of \NPM,
{\it cf.} Eq.(23),
\beq C_q(M)=M^q\la\omega^q\ra^\nu=M^q\e^{\nu\ln\la\omega^q\ra} =
M^{\phi_q}, \eeq 
with $M=2^\nu=\e^{\nu\ln 2}$, $\phi_q=q+\ln\la\omega^q\ra/\ln 2$.

It is easy to see using Eq.(10) that the \NPM\ $C_q$ can be
extracted from experimental data through the normalized factorial
moments (\NFM) 
\beq F_q(M) = M^{q-1}\sum_{m=1}^M
\frac{\la n_m(n_m-1)\cdots(n_m-q+1)\ra}{\la n\ra^q},  \eeq 
The anomalous scaling of \NFM, usually referred to as
intermittency~\cite{BP}\cite{DDK}, has been observed in
hadron-hadron~\cite{NA22SF}\cite{NA27SF} and \EE~\cite{CGDatong}
experiments, providing a first signal for the non-linear fractal
property of strong interaction dynamics.

The \NPM\ defined in Eq.(23) can be viewed as the sample average
of ``event-probability moment'' (\EPM) 
\beq C_q^{(\rm e)}(M) = M^{-1}\sum_{m=1}^M (Mp_m)^q.\eeq
It is natural to consider the distribution of \EPM\ itself instead
of only its average~\cite{HWA}. This distribution can be
characterized by the $p$th order event-space moment $C_{q,p}$ of
the $q$th order event-moment $C_q^{(\rm e)}$  
\beq C_{q,p}(M)= \f\la \f( C_q^{(\rm e)}(M) \g)^p \g\ra =
\f\la \f(M^{-1}\sum_{m=1}^M (Mp_m)^q\g)^p\g\ra  \eeq 
and the corresponding normalized moment 
\beq C_{q,p}^{(\rm norm)}= C_{q,p}/(C_{q,1})^p. \eeq 
In order to measure $C_{q,p}$ in real experiments, the $p$th order
event-space moment of the $q$th order event-factorial-moment 
\beq F_{q,p}=\f\la \f( 
M^{q-1} \sum_{m=1}^M \frac{n_m(n_m-1) \cdots (n_M-q+1)}
{\la n\ra^q} \g)^p \g\ra \eeq 
has been proposed~\cite{HWA}. The scaling property of $F_{q,p}$
with the increase of $M$ is referred to as erraticity~\cite{HWA}
and has been observed in experiments~\cite{NA27err}\cite{NA22err}.
However, when $p\neq 1$, $F_{q,p}$ contains statistical
fluctuations~\cite{ZGKX} and is unequal to $C_{q,p}$, despite of
the factorial moments apparently used. The observed erraticity
phenomena have been shown to be dominated by statistical
fluctuations and the dynamical effect is
hidden~\cite{ZGKX}\cite{Staterr}.

In order to be able to observe the dynamical ``erraticity'', the
statistical fluctuation should be eliminated first. The
elimination of (Poissonian) statistical fluctuation for the case
$p=1$ is straightforward using Eq.(10). The result is just Eq.(31)
with $p=1$, or Eq.(27). Let us consider the elimination of
statistical fluctuations for the cases $q=2$, $p=2,3$. We have 
 \bea C_{2,2}&=&\f\la \f( M \sum_{m=1}^M p_m^2\g)^2\g\ra =
M^2\f\la \sum_{m=1}^M p_m^4 + \sum_{m\neq m'}^M
p_m^2p_{m'}^2\g\ra ,  \\ 
 C_{2,3}&=&\f\la \f( M \sum_{m=1}^M p_m^2\g)^3\g\ra =
M^3\f\la \sum_{m=1}^M p_m^6 + 3\sum_{m\neq m'}^M p_m^2p_{m'}^4+
 \sum_{m\neq m'\neq m''}^M p_m^2p_{m'}^2p_{m''}^2 \g\ra. \ \ \eea 
Define 
\beq G_{2,2}= M^2\f\la \sum_{m=1}^M \frac{n_m \cdots (n_m-3)}{\la
n\ra^4} \g\ra + M^2 \f\la \sum_{m\neq m'}^M
\frac{n_m(n_m-1)n_{m'}(n_m'-1)}{\la n\ra ^4}\g\ra , \eeq 
\bea G_{2,3} \nonumber &=& M^3\f\la \sum_{m=1}^M \frac{n_m\cdots
(n_m-5)}{\la n \ra^6}\g\ra + 3 M^3 \f\la \sum_{m\neq m'}^M \frac{
n_m(n_m-1)n_{m'}\cdots (n_{m'}-3)}{\la n\ra ^6}\g\ra \\
 &+& M^3 \f\la \sum_{m\neq m'\neq m''}^M
 \frac{n_m(n_m-1)n_{m'}(n_{m'}-1)n_{m''}(n_{m''}-1)}
{\la n\ra^6} \g\ra . \eea 
Then, utilizing Eq.(10), it is ready to show that
$G_{2,2}=C_{2,2}$, $G_{2,3}=C_{2,3}$, provided the statistical
fluctuations are Poissonian. This means that using $G_{2,2}$,
$G_{2,3}$ instead of $F_{2,2}$, $F_{2,3}$ to measure the dynamical
moments $C_{2,2}$, $C_{2,3}$, the (Poissonian) statistical
fluctuations are eliminated and the real dynamical ``erraticity''
is observed. This method can be extended to any positive integer
orders $q$ and $p$.

To illustrate the above result by Monte Carlo simulation, the
random cascading $\alpha$ model described above is used with a
Gaussian distributed parameter $\alpha$ (mean=0.5, width=0.15).
Particles are put into each event according to Poisson
distribution with $\la n \ra =6$. The results from 2,500,000
generated events are plotted in Fig.2. The coincidence of
$G_{q,p}^{\rm norm}$ and $C_{q,p}^{\rm norm}$ is remarkable. It is
to be contrasted with the strongly upward bending curves of
ln$F_{q,p}^{\rm norm}$ {\it vs.} ln$M$. This shows that the
strongly upward-bending behavior of ln$F_{q,p}^{\rm norm}$ {\it
vs.} ln$M$ for positive $p$ which is typical in the conventional
erraticity analysis~\cite{HWA}\cite{NA27err}\cite{NA22err} is due
to statistical fluctuations. The reason is: when the partition
number $M$ increases, the average multiplicity per bin becomes
smaller and smaller and the statistical fluctuations get stronger
and stronger. After eliminating the statistical fluctuations, the
ln$G_{q,p}^{\rm norm}$ {\it vs}. ln$M$ curve coincides with
ln$C_{q,p}^{\rm norm}$ {\it vs}. ln$M$ and recovers the anomalous
scaling property of the event dynamics (dynamical erraticity) in
the model.

\begin{center}
 \begin{figure}
 \includegraphics[width=12cm]{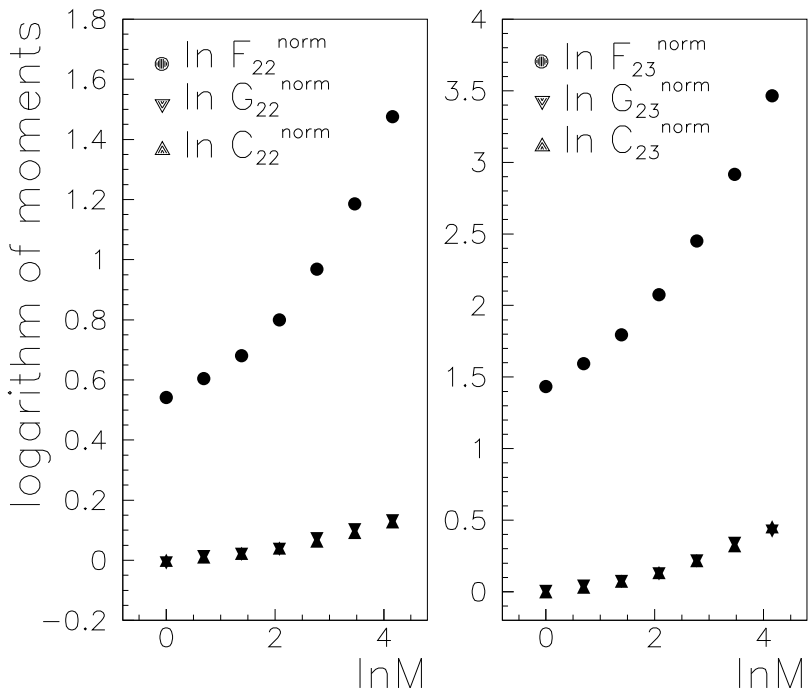}
 \caption{\label{fig:cqp} Scaling property of event-space moments}
 \end{figure}
 \end{center}

\vs 5mm {\bf V \ Conclusions} \vs 1mm

In this paper the problem of eliminating the random noise in
event-by-event analysis is considered. New moments $G_p$
(for continuous distribution), and $G_{q,p}$ (for anomalous
distribution) are proposed, which are experimentally measurable
and can be used to eliminate the Poissonian type statistical fluctuations
and recover the dynamical moments $C_p$ and $C_{q,p}$.

For a comparison of different methods we notice that most of the
measures proposed in the market for the dynamical fluctuation of
transverse momentum are based on a subtraction procedure, {\it
i.e.} to subtract the variance of $\bar{\pt}$ or a quantity
related to it, that will be expected from pure statistical
fluctuation, from the same quantity obtained in experiment. In the
present paper we take another approach, {\it i.e.} not to evaluate
the effect of pure statistical fluctuation and subtract it, but to
eliminate the statistical fluctuations directly and quantitatively
from the experimental data and recover the dynamical
$\sigma^2_{\bar{\pt}}$ under the unique assumption that the
statistical fluctuation is Poissonian, {\it i.e.} due to
uncorrelated random particle emission.

The difference between these two approaches is two-fold.

Firstly, in the subtraction method how to get the ``pure
statistical'' variance from experimental data is a big problem.
The mixing-event method~\cite{PHENIX} can in principle get the
statistical variance, but the accuracy depends on the
mixing-procedure. The method proposed in Ref.~\cite{Voloshin}
relies on the equality $\sigma^2_{{\bar{\pt}}\rm
stat}=\sigma^2_{\pt \rm incl}/\la n\ra$, {\it cf.} Eq.(6), but
this equality holds only for a pure statistical system without any
dynamical fluctuation. In experimental data sample there exist
simultaneously dynamical and statistical fluctuations, and the
statistical variance $\sigma^{2{\rm data}}_{{\bar{\pt}}\rm stat}$
included in the data sample is unequal to the inclusive variance
of the data sample over average multiplicity: 
\beq  \sigma^{2{\rm data}}_{{\bar{\pt}}\rm stat}\neq \sigma^{2{\rm
data}}_{\pt \rm incl}/\la n\ra. \eeq 
The same holds also for the $\Phi_{\pt}$ method~\cite{marek}.

Secondly, the variance $\sigma^2_{\bar{\pt}}$ contains the
$\bar\pt$ moment only up to second order, which only gives the
width of the distribution of $\bar{\pt}$ in the event space. Using
the method proposed in the present paper the $\bar\pt$ moment of
any integer order can be extracted and the dynamical distribution
of $\bar{\pt}$ in the event space is portrayed in much more
detail.

In the anomalous case, the difficulty coming from the dominance of
statistical fluctuation in the theoretical and experimental
studies of erraticity up to now can be overcome and the dynamical
erraticity, if exists, can be extracted using the method proposed
in the present paper.

Applying the proposed method to real experimental data is highly
recommended.

{\bf Acknowledgement} The authors thank Wu Yuanfang and Liu Feng for helpful
discussions and comments.

\ed
\begin{thebibliography}{9}
\bibitem{note1} In experiment the smallest $\delta_m$ is limited
by the resolution of $x$.
\bibitem{note2} Through out the present paper the symbol $\la x\ra$ is used to
denote the sample average of $x$, no matter $x$ is an event
variable or an inclusive variable, and the symbol $\bar x$ is
reserved for the event average of $x$.
\end{thebibliography}
